\documentclass[%
 aip,
cp,  
 amsmath,amssymb,
 reprint,%
]{revtex4-2}

\usepackage{graphicx}
\usepackage{dcolumn}
\usepackage{bm}
\usepackage{sidecap}

\usepackage[utf8]{inputenc}
\usepackage[T1]{fontenc}
\usepackage{mathptmx} 

\begin{document}
\title{Not so presto? Can outer hair cells be sluggish?}

\author{Kuni H. Iwasa} 
 \email[Corresponding author: ]{kuni.iwasa@gmail.com}
\affiliation{NIDCD - National Institutes of Health Bldg.\ 35A Rm 1F121A,
  Bethesda Maryland 20892 USA
}

\date{\today} 

\begin{abstract}
Prestin (SLC26A5), a protein essential for the sensitivity of the mammalian ear, was so named from \emph{presto}. The assumption was that this membrane protein supports fast movement of outer hair cells (OHCs) that matches the mammalian hearing range,  up to 20 kHz in general and beyond, depending on the species.  \emph{In vitro} data from isolated OHCs appeared to be consistent with such frequencies. However, some recent reports cast doubts on this assumption, suggesting that the intrinsic transition rates of this protein are much lower, about 3 kHz for guinea pigs, not covering the auditory frequency range of the animal. Recent \emph{in vivo} data also show that the amplitude of OHC motion rolls off well below the best frequency of the location. The present report examines whether or not these recent observations are compatible with the physiological function of OHCs by using simple piezoelectric models.
\end{abstract}

\maketitle

\section{Introduction}\label{sec:intro}
The significance of prestin for the sensitivity and frequency selectivity of the mammalian ear has been well established \cite{Dallos2008}. However, the detailed mechanism, with which this piezoelectric membrane protein plays its physiological role, remains not as clear. The frequency range that this protein is capable of responding is a critical issue.

Earlier \emph{in vitro} studies on isolated OHCs confirmed that their fast motile response is based on piezoelectricity \cite{i1993,ga1994,doi2002}. Force generation under quasi-isometric condition was shown to have flat frequency dependence up to 60 kHz \cite{fhg1999}. The characteristic frequency of the power spectrum of membrane current due to prestin was about 40 kHz \cite{Dong2000}. Those frequencies were considered to be lower bounds imposed by experimental conditions.

However, recent reports on OHCs in \emph{in vitro} \cite{Santos-Sacchi2018,SantosSacchi2019} appear to contradict these earlier observations. 
In addition, \emph{in vivo} data obtained with optical coherence tomography (OCT) \cite{Vavakou2019}  have been interpreted as evidence against fast motile response of OHCs.

The present report examines the implications of these two kinds of reports by asking two questions. First, can OHCs counteract local viscous drag incurred by the movement of the organ of Corti in the cochlea if gating rates of prestin are as low as these recent reports suggest?  At the basal end, where the traveling wave initiates, viscous drag must be counteracted by OHCs. This condition determines the frequency limit of the ear. 

Second, should the movement of OHCs large at their best frequencies \emph{in vivo}? The first issue is addressed by evaluating power generation by OHCs driven by prestin with finite transition rates. The second issue is addressed using a simple model system, which consists of two harmonic oscillators, one of which has a driver and the other with a damper.

\section{1. Transition rates}\label{sec:intrinsic}
First, let us derive the equation of motion for a cell, which is driven by a motile molecule with finite intrinsic transition rates. An earlier version has been published as Appendix to Santos-Sacchi et al.\ \cite{SantosSacchi2019}. 

\subsection{Rate equation}
Consider a membrane molecule with two discrete conformational states C$_0$ and C$_1$ and let the transition rates $k_+$ and $k_-$ between them, with $k_+$ from C$_0$ and C$_1$, and $k_-$ the opposite, i.e.
\begin{align}\nonumber
&\;k_+ \\
 C_0 \quad &\rightleftharpoons\quad C_1 \\ \nonumber
 &\;k_-
\end{align}

Let $P_1$ be the probability that the molecule in state C$_1$, which elongate the cell.
Suppose charge transfer $q (>0)$ is associated with a change $a(>0)$ in the length of the cell. Then $P_1$ satisfies
\begin{align}\label{eq:P1qa}
\frac{P_1}{1-P_1}=\frac{k_+}{k_-}=\exp[\beta [q(V-V_0)+aF]],
\end{align}
where $F$ is the axial force applied on the cell. The transition rates $k_+$ and $k_-$ can be expressed as
\begin{subequations}\label{eq:kpmF}
\begin{align}
k_+&=\exp\left[\alpha\beta  [q(V-V_0)+aF]\right], \\
k_-&=\exp\left[(-1+\alpha)\beta  [q(V-V_0)+aF]\right],
\end{align}
\end{subequations}
by introducing a parameter $\alpha$.
For the rest of the present paper, the dependence on the value of the parameter $\alpha$ does not appear except for $\omega_g(=k_++k_-)$. 

The time dependence of $P_1$ can be expressed by the rate equation
\begin{align}
\frac {d}{dt}P_1=k_+-(k_++k_-)P_1.
\end{align}

Now assume that the voltage $V$  consists of a constant term $\overline V$ and a small sinusoidal component with amplitude $v$ , i.e. 
$
V=\overline V+v\exp[i\omega t], 
$
where $\omega$ is the angular frequency and $i=\sqrt{-1}$. Then the transition rates are time dependent due to the voltage dependence, satisfying
\begin{align}\label{eq:kratio}
\frac{k_+}{k_-}=\frac{\overline{k}_+}{\overline{k}_-}(1-\beta qv\exp[i\omega t]),
\end{align}
where $\overline{k}_+$ and $\overline{k}_-$ are time independent, and small amplitude $v$ implies $\beta qv \ll 1$. A set of $k_+$ and $k_-$ that satisfies Eq.\ \ref{eq:kratio} can be expressed
\begin{subequations}
\begin{align}\label{eq:kp}
k_+&=\overline{k}_+(1-\alpha \beta qv\exp[i\omega t]),\\ \label{eq:km}
k_-&=\overline{k}_-\{1-(1-\alpha) \beta qv\exp[i\omega t]\}.
\end{align}
\end{subequations}

If we express $P_1=\overline{P}_1+p_1\exp[i\omega t]$, for the 0th and 1st order terms  we have, respectively \cite{i1997}
\begin{subequations}
\begin{align}
\overline P_1&=\frac{\overline{k}_+}{\overline k_++\overline k_-},\\ \label{eq:p_1}
p_1 &=-\frac{\overline{k}_+\overline{k}_-}{\overline k_++\overline k_-}\cdot\frac{\beta qv}{i\omega+\overline{k}_++\overline{k}_-}.
\end{align}
\end{subequations}
This simple result can be obtained in the case $\alpha=\overline k_-/(\overline k_++\overline k_-)$. 

By using the relationship $\overline P_1=\overline k_+/(\overline k_++\overline k_-)$, Eq.\ \ref{eq:p_1} turns into
\begin{align}\label{eq:p-1}
 \quad  p_1=\frac{\beta q \overline P_\pm}{1+i\omega/\omega_g}\cdot v,
\end{align}
with $\overline P_\pm=\overline P_1(1-\overline P)$ and $\omega_g=\overline k_++\overline k_-$.
Here, the average values and the amplitudes of sinusoidal components are represented by notations similar to the voltage. Notice that Eq.\ \ref{eq:p-1} satisfies the Boltzmann distribution. That means the frequency dependence is determined only by the transition rates and unaffected by mechanical load or by mechanical relaxation.

\begin{SCfigure}[1]
\includegraphics[width=4.15cm]{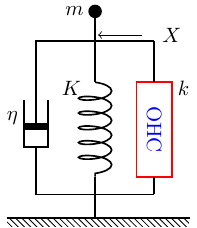} 
\caption{
A schematic representation of a cell with mechanical load. The stiffness of the cell is $k$, the stiffness of the external elastic load is $K$, the inertial load is $m$, and drag coefficient is $\eta$. The state of the outer hair cell (OHC) is described by $P$. The mechanical displacement of the cell is represented by $X$. The properties of OHC cannot easily be rendered as a simple combination of elastic- and displacement elements, even though such attempts have been made earlier \cite{Iwasa2016,Iwasa2017,Iwasa2021}.}
\label{fig:simple}
\end{SCfigure}

\subsection{Equation of motion}\label{subsec:EoM}
Consider a system, where an OHC is associated with mass $m$, drag coefficient $\eta$, and external elastic load $K$ (Fig.\ \ref{fig:simple}). Assume that the material stiffness of the OHC is $k$ \footnote{The actual axial stiffness is reduced by conformational transitions of the motile elements \cite{i2000}. This effect is analogous to ``gating compliance'' \cite{hh1988}}. The sum of inertial force and drag is balanced with  elastic force $F(t)$. The displacement $X(t)$ satisfies the equation
\begin{align}\label{eq:XF}
 \left(m\frac {d^2}{dt^2}+\eta\frac{d}{dt}\right)X(t)=F(t).
\end{align}
If we are interested in the response to the stimulus with an angular frequency $\omega$, the equation of motion Eq.\ \ref{eq:XF} can be expressed as
\begin{align}\label{eq:xf}
( -\omega^2 m+i\eta\omega)x=f(\omega),
\end{align}
where $x$ and $f$ are, respectively, the amplitudes of displacement $X$ and force $F$ with the frequency $\omega$.

If the force $F$ in Eq.\ \ref{eq:kpmF} is due to the external load alone, we have $F=-KX$ and $f=-Kx$, which are the familiar equations of motion. In such a case, the equilibrium condition is $X=0$ and $x=0$. In our system, however, movement is driven by a deviation from the Boltzmann distribution. If the displacement can respond instantaneously to voltage changes, the system goes from one equilibrium to another.

Note that the quantity $p_\infty$ is similar to $p_1$ in Eq.\ \ref{eq:p-1} in that it satisfies the Boltzmann distribution. However, it depends on both $v$ and $p$ because the energy term has both electrical and mechanical terms as expressed by Eqs.\ \ref{eq:kpmF}. If the difference between $p_\infty$ and $p$ is small, the driving force can be proportional to the difference $p_\infty-p$. Thus, the driving force can be expressed
\begin{align}
f=k\cdot aN(p_\infty-p). 
\end{align}

The presence of external elastic load $K$ makes the displacement $x$ expressed by $x=aNp\cdot k/(k+K)$, where $n$ is the number of motile elements. By choosing $p$ as the variable of the equation, we have
\begin{align}\label{eq:eom_fre}
\left[-\omega^2 m+i \eta\omega\right]p&=(k+K)(p_\infty-p),
\end{align}
which has a familiar form for the equation of motion.

For a given set of values for $v$ and $p$, the quantity $p$ tends to move to the value that satisfies equilibrium condition, which is given by
\begin{align}\label{eq:ap1v}
p_\infty=\beta \overline P_\pm (qv+a^2N\tilde Kp).
\end{align}
with $\tilde K=kK/(k+K)$ because voltage changes affects length changes as well as charge transfer \cite{Iwasa2016}. 

With finite transition rates of prestin, the response is to an earlier state of $p_\infty$. The time delay is characterized by $\omega_g$ as expressed by Eq. \ref{eq:p_1}. By introducing the explicit form of $p_1$, this equation turns into
\begin{align}\label{eq:p1}
  [-(\omega/\omega_r)^2+i \omega/\omega_\eta+1+\delta^2]p=\frac{\beta \overline P_\pm}{1+i\omega/\omega_g}\cdot qv,
\end{align}
\noindent where $\omega_r^2 =m/(k+K)$, $\omega_\eta =\eta/(k+K)$, and $\delta^2=\beta \overline P_\pm na^2\tilde K/(1+i\omega/\omega_g)$. The quantity $\delta^2$ has only a minor effect on $p$. 
Notice if we let $\omega_g\rightarrow\infty$, factor $1/(1+i\omega/\omega_g)$ turns into unity, we obtain the equation  \cite{Iwasa2017} for infinitely fast gating.

In a special case of $m=K=0$, this equation turns into
\begin{align}
  (1+i \omega/\omega_\eta)(1+i\omega/\omega_g)p=\beta \overline P_\pm\cdot qv,
\end{align}
showing low-pass behavior with two time constants $1/\omega_g$ and $1/\omega_\eta$.

\subsection{Power output}\label{subsec:power}
Power output $W(\omega)$ by OHCs can be evaluated by calculating the work done against viscous drag because the work against elastic load is recovered after a cycle. The mean power output $\langle W(\omega)\rangle$ is given by 
$ \langle W(\omega)\rangle=(\eta/2)|\omega aNp/(k+K)|^2$.

\begin{SCfigure}[1] 
\includegraphics[width=0.3\textwidth]{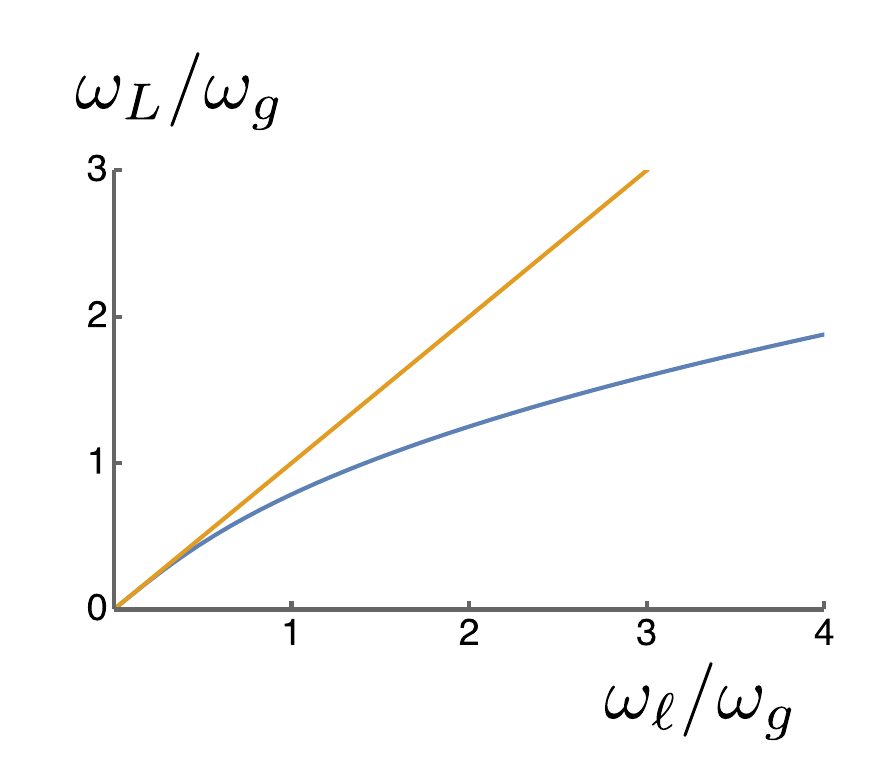}
\caption{Limiting frequency $\omega_L$ for prestin with gating frequency $\omega_g$ against that of $\omega_\ell$, for infinitely fast gating. Here, $\omega_L/\omega_g$, the limiting frequency normalized with respect to the gating frequency  is plotted against $\omega_\ell/\omega_g$ (blue). Only in the limit of large $\omega_g$, i.e. for small $\omega_\ell/\omega_g$ and$\omega_L/\omega_g$ , does $\omega_L$ approach $\omega_\ell$, the straight orange line.}
\label{fig:freqL}
\end{SCfigure}

Eqs.\ \ref{eq:p1}, with the observation that $\delta^2$ has a relatively minor effect on $p$, indicates that the mean power generation $\langle W_\infty(\omega)\rangle$ for the special case of infinitely fast gating differs from $\langle W(\omega)\rangle$ only in the absence of the attenuation factor $1/[1+(\omega/\omega_g)^2]$ if $v$ remains the same. However, the factor $1/(1+i\omega/\omega_g)$ reduces $v$ because it reduces movement of the motile element. For this reason, we may put
\begin{align}
 \langle W(\omega)\rangle<\langle W_\infty(\omega)\rangle/[1+(\omega/\omega_g)^2].
\end{align}

Recall how an optimal limiting frequency was determined for the case, where gating is infinitely fast. In the presence of inertia, power generation has a peak value $P_\infty^\mathrm{max}$ due to piezoelectric resonance.  A limiting frequency $\omega_\ell$ for that case is obtained from  
$P_\infty^\mathrm{max}=\mu\omega_\ell^2,$
equating the maximal power production with viscous loss. Here $\mu$ is proportional to viscous coefficient $\eta$.

If we can assume that the resonance peak is sharp, the corresponding limiting frequency $\omega_{L}$ for a finite gating frequency $\omega_g$ can be approximated by
$P_\infty^\mathrm{max}/[1+(\omega_L/\omega_g)^2] < \mu\omega_L^2.$


The combination of these two equations leads to
\begin{align}\label{eq:limit}
(\omega_L/\omega_g)^2<\frac 1 2\left( \sqrt{1+4(\omega_\ell/\omega_g)^2}-1\right).
\end{align}
\noindent For $\omega_\ell=2\pi \times 10$ kHz \cite{Iwasa2017} and $\omega_g=2\pi\times 3$ kHz \cite{SantosSacchi2019}, $\omega_L/\omega_g<1.1$ (See Fig. \ref{fig:freqL}). The limiting frequency is 3.3 kHz, not much higher than the gating frequency.

The assumption that led to the inequality (\ref{eq:limit}) is not always satisfied because the resonance peak disappears as the gating frequency decreases. In such cases, however, the inequality still holds even though the real frequency limit is lower than indicated because power production is lower without resonance peak.

\section{2. Coupled oscillators}\label{sec:coupled}
The evaluation of maximum power output as mentioned above is based on the assumption that OHCs operate at near resonance frequencies. That assumption in turn requires multiple modes of motion in the organ of Corti because of the mismatch in stiffness between OHCs and the basilar membrane \cite{Iwasa2017}. In the following, a coupled oscillator is used as a model illustrate this issue.

Here we assume an OHC is associated with an oscillator, which is weakly coupled with another oscillator. It would be useful to examine a system of interacting coupled harmonic oscillators explicitly due to its simplicity, even though the cochlea is not a linear system. The following analysis can be justified only for low input level.

\subsection{Coupled harmonic oscillators}\label{subsec:harmonic}
A textbook example of coupled harmonic oscillators are connected by an elastic element. Typically described by
\begin{subequations}\label{eq:DO_DO}
\begin{align}
\left(M_1 \frac{d^2}{dt^2}+\eta_1 \frac{d}{dt}+K_1\right)X_1&=K_c(X_2-X_1)\\
\left(M_2 \frac{d^2}{dt^2}+\eta_2 \frac{d}{dt}+K_2\right)X_2&=K_c(X_1-X_2)+F_2(t),
\end{align}
\end{subequations}
where $M_j$ is the mass, $\eta_j$ drag coefficient, and $K_j$ the spring of each oscillator (for j=1,2). $K_c$ is the stiffness of the spring that connects two oscillators. $F_2(t)$ is the force that stimulates the system.  The classical analyses  of such coupled oscillators are available \cite{Morse_Ingard,Mercer1971}. Despite the simplicity of the idea, the frequency dependence of the energy transfer between the two oscillators is rather complex \cite{Mercer1971}.

\subsubsection{Amplification with negative drag}\label{subsec:neg_drag}
The simplest introduction of active or amplifying mechanism to the system of interacting harmonic oscillator would be to flip the sign of the drag term in one of the oscillators. 

Let us assume that the first one with displacement $x$ is an amplifying oscillator (AO), i.e. $\eta_1<0$, and the second one with displacement $y$ is a dissipating oscillator (DO) and stimulus is applied to the DO. Let the stimulus have a sinusoidal waveform with at an angular frequency $\omega$. Then $F_2=f_2\exp[i\omega t]$ and the amplitudes of AO and DO are, respectively, $x$ and $y$.
 
The equation of motion can be expressed in the form
\begin{subequations}
\begin{align}\label{eq:asymmetric1}
[-(\overline\omega/\overline\omega_1)^2-i\overline\omega/\overline\omega_{a}+1+cs]x-cs y&=0,\\
-cx+[-\overline\omega^2+i\overline\omega/\overline\omega_{\eta}+1+c]y&=f,
\end{align}
\end{subequations}
where frequencies are normalized with respect to the resonance frequency $\omega_2$ (where $\omega_2^2=M_2/K_2$) of the DO, i.e.\ $\overline\omega=\omega/\omega_2$, $\overline\omega_\eta=\omega_\eta/\omega_2$, and $\overline\omega_a=\omega_a/\omega_2(>0)$. The negative sign on the term involving $\overline\omega_a$ indicates negative drag that amplifies the oscillation of the system. 

The parameters introduced are $f=f_2/K_2$, $c(=\!K_c/K_2)<1$ and $s(=\!K_2/K_1)<1$, because it is reasonable to assume the stiffness of the basilar membrane is larger than the elastic load of OHCs. Notice that the ratio $x/y$ is larger than $X_1/X_2$ because $s\;X_1/X_2=x/y$.

\subsubsection{OHC as amplifier}
Eq. \ref{eq:eom_fre} for an OHC with respect to the variable $p$ can be rewritten again by using the displacement $x$, which is defined by $aNpk/(k+K)$, as
\begin{align}
(-\omega^2m+i\eta\omega+1)x=(k+K)aNkp_\infty,
\end{align}
assuming the gating of prestin depends on the mechanical factor alone. Here, the component of the receptor potential $v$ of the stimulation frequency $\omega$ is determined by
$-i_0\hat{r}=(\sigma+i\omega C_0)v-i\omega Nqp$.
Here $i_0=(e_{ec}-e_K)/(\overline R_a+R_m)$, where $e_{ec}$ is the endocochlear potential,  $e_K$ the resting potential of the OHC, $\overline R_a$ the resting level of hair bundle resistance, $R_m$ the basolateral resistance, and $C_0$ the regular capacitance. 

The quantity $\hat{r}$ is the relative change of the hair bundle conductance. If $\hat{r}$ depends on $x$, the equations of motion can be expressed as
\begin{subequations}
\begin{align}
[-\overline\omega^2-iA/\overline\omega+1+cs-B]x-csy=0,\\
-cx+[-\overline\omega^2-i\overline\omega/\overline\omega_\eta+1+c]y=f,
\end{align}
\label{eq:x-drive}
\end{subequations}
where $A$ and $B$ are determined by the operating point. This set of equations is similar to the one with anti-drag amplifier, Eq.\ \ref{eq:asymmetric1} even though that the frequency dependence of the amplifier term is different.

\begin{SCfigure}[0.7]
\includegraphics[width=10cm]{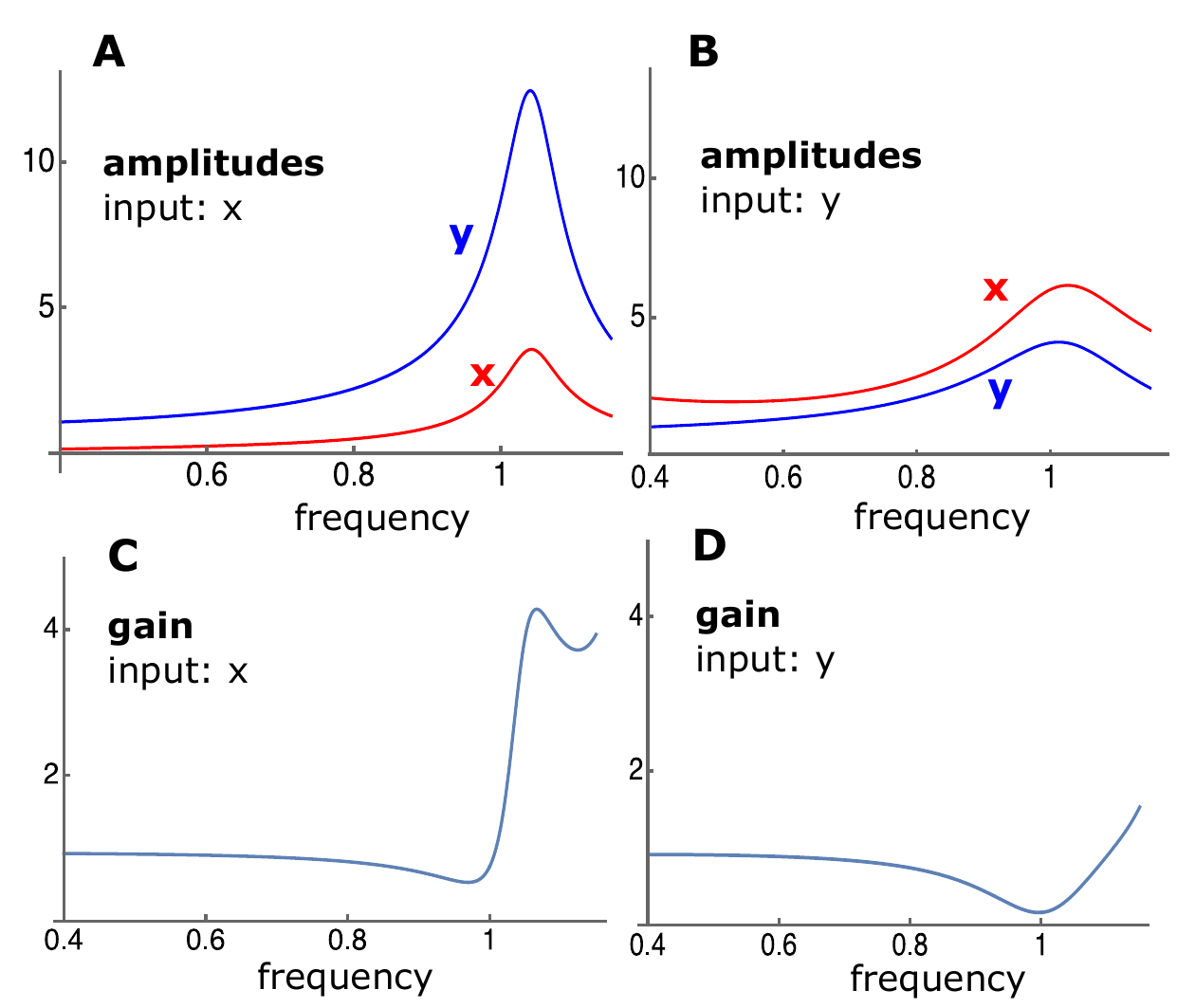} 
\caption{Amplitudes and gain plotted against $\omega/\omega_r$, the frequency normalized to the resonance frequency. Here both oscillators are assumed to have the same resonance frequency. The amplitudes are normalized to external force $f$. Gain is the square of ratio of the amplitude $y$ while OHC is engaged to the amplitude $y$ while OHC is not engaged. Left (A and C): input to the hair bundle is x. Right (B and D): input to the hair bundle is y. Gain is the ratio of the squared amplitude of $y$ relative to the condition, where OHC is turned off. The reduction of the gain near reduced frequency unity is associated to a positive peak shift due to OHC, i.e. the effect of the parameter $B$ in the equations. Parameter values are, $\omega_1 = 1, \; c = 0.1, \; k = 5, \;\overline\omega_\eta = 10, \;s = 0.01, \;B = 0.57, \; A = 1.5$.}
\label{fig:amp_gain}
\end{SCfigure}

If the relative change $\hat r$ of hair bundle resistance depends on $y$, the other oscillator, the set of equations take another form:
\begin{subequations}
\begin{align}
[-\overline\omega^2 +1+ck-B]x-[iA/\overline\omega+cs]y=0,\\
-cx+[-\overline\omega^2+i\overline\omega/\overline\omega_\eta+1+c]y=f.
\end{align}
\label{eq:y-drive}
\end{subequations}
Some of numerical examinations of these equations are shown in Fig. \ref{fig:amp_gain}. 
\subsubsection{Numerical examination}
Having identified variable $x$ as associated with OHC, and $y$ with the BM to derive Eqs.\ \ref{eq:x-drive} and \ref{eq:y-drive}, it could be more convenient to call the two oscillators as OHC-associated and BM-associated rather than amplifying and dissipating oscillators.

Fig.\ \ref{fig:amp_gain} shows that the sensitivity of the hair bundle to OHC displacement is critical for the amplifier gain (Fig.\ \ref{fig:amp_gain}C) near the resonance frequency and that the amplitude of OHC displacement could be much smaller than that of the BM. This result appears consistent with experimental observations \cite{Vavakou2019}. However, the small amplitude of OHC movement does not increase at lower frequencies and it is not consistent with the reported experimental observation \cite{Vavakou2019}.

If the hair bundle is driven by the BM, the amplitude of OHC displacement is larger. At the same time the gain is smaller. This condition could be regarded as similar to the experimental observation at low frequencies. Thus, the experimental observations can be explained if we can assume that the hair bundle is stimulated by the BM at low frequencies and by OHC displacement near the characteristic frequency.

This idea is incompatible with the assumption that hair bundle displacement is a weighted sum of BM displacement and OHC displacement throughout the frequency range.  That is because the much smaller amplitude of the OHC needs to dominate hair bundle stimulation to make the amplitude of the BM larger.

\section{Conclusions}\label{sec:conclude}

The intrinsic transitions of prestin, which is internal conformational rearrangement of the molecule, must be fast enough to exceed the upper bound of the auditory range for electromotility to counteract local drag. Even though energy can flow laterally along the cochlea, a large gap between the intrinsic gating frequency and the best frequency of the location makes it unlikely for OHCs function as the cochlear amplifier. For these reasons, low frequency gating of prestin in recently reports \cite{Santos-Sacchi2018,SantosSacchi2019}, which contradicts older reports \cite{fhg1999,Dong2000}, is not compatible the OCT data \cite{Vavakou2019} because low frequency gating is incapable with amplifying effect of OHCs at best frequencies.

The organ of Corti is an anisotropic, heterogeneous-three dimensional object. The present analysis, based on a set of simple coupled oscillators, is intended to provide a simplest possible model for such a complicated system, ignoring nonlinearity. characteristic to the cochlea. With these reservations, the following conclusions can be drawn.
OHCs can function as cochlear amplifier, even if the amplitude of OHC length changes is much less than the amplitude of basilar membrane movement, consistent with OCT data \cite{Vavakou2019}. 

The model presented by itself cannot explain the amplitude roll-off of OHC movement as observed with the OCT  \cite{Vavakou2019}. However, the present analysis suggests that such a roll-off can be explained as the consequence of mode changes. In other words, the observed frequency dependence of the mode of motion of the organ of Corti \cite{Gao2014,Cooper2018} could indicate the mode of motion at low frequencies is  different from that of at the best frequencies.  Verifying this expectation would require a detailed analysis of modes of motion \emph{in vivo} and their dependence on the frequency.

\section{Acknowledgment}
This work was partially supported by the Intramural Program of NIDCD, NIH.
%

\end{document}